\documentclass[aps,amsfonts,amsmath,prd,preprint,nofootinbib]{revtex4}
\newcommand{\beq}{\begin{equation}}
\newcommand{\eeq}{\end{equation}}

\usepackage{epsfig,bbm,cancel}
\usepackage[breaklinks=true]{hyperref}
\newcommand{\mc}[1]{\mathcal{#1}}

\begin{document}

\title{Some exact BPS solutions for exotic vortices and monopoles}

\author{Handhika S. Ramadhan}
\email{hramad@ui.ac.id}
\affiliation{Departemen Fisika, FMIPA, Universitas Indonesia, Depok 16424, Indonesia. }

\def\changenote#1{\footnote{\bf #1}}

\begin{abstract}

We present several analytical solutions of BPS vortices and monopoles in the generalized Abelian Maxwell-Higgs and Yang-Mills-Higgs theories, respectively. These models have recently been extensively studied and several exact solutions have already been obtained in~\cite{Casana:2014qfa, Casana:2013lna}. In each theory, the dynamics is controlled by the additional two positive scalar-field-dependent functions,  $f(|\phi|)$ and $w(|\phi|)$. For the case of vortices, we work in the ordinary symmetry-breaking Higgs potential, while for the case of monopoles we have the ordinary condition of the Prasad-Sommerfield limit. Our results generalize that of exact solutions found previously. We also present solutions for BPS vortices with higher winding number. These solutions suffer from the condition that $w(|\phi|)$ has negative value at some finite range of $r$, but we argue that since it satisfies the weaker positive-value conditions then the corresponding energy density is still positive-definite and, thus, they are acceptable BPS solutions.

\end{abstract}

\maketitle
\thispagestyle{empty}
\setcounter{page}{1}

\section{Introduction}

Bogomol'nyi-Prasad-Sommerfield (BPS) equations are a set of first-order equations in field theory that minimizes the action and saturates the lower bound of the corresponding static energy. In his seminal paper, Bogomol'nyi~\cite{Bogomolny:1975de} shows that by completing the squares of the corresponding Hamiltonian the second-order Euler-Lagrange equations of motion for domain walls, strings, or monopoles can be reduced to first-order\footnote{This is not always an easy task, considering the  complicacy of the Lagrangian. Recently, several alternative formalisms have been proposed to derive the BPS equations~\cite{Atmaja:2014fha, Atmaja:2015lia, Atmaja:2015umo}. These formalisms do not start from the Hamiltonian, but rely directly on the Euler-Lagrange instead. The known BPS equations can be shown to be the result of taking separation-of-variable ansatz for the corresponding auxiliary functions.}, and their static energies are global minima; thus stable. In the context of monopoles, Bogomol'nyi method provides a physical explanation of the exact Prasad-Sommerfield solutions~\cite{Prasad:1975kr}.

In the last few years, there has been an extensive discussions on defects with noncanonical kinetic terms~\cite{Babichev:2006cy, Babichev:2007tn, sarangi, Babichev:2008qv, pavlovski, ramadhan1, ramadhan2}. These noncanonical defects gain interest, partly, because they can evade Derrick's constaint for the existence of solitons~\cite{Derrick:1964ww}, but also on a more applicative level can be used as effective models in cosmology; for example in the study of: inflationary phase of the present universe using the $k$-essence models~\cite{Babichev:2007dw, ArmendarizPicon:1999rj}, dark matter~\cite{ArmendarizPicon:2005nz}, seeds for structure formation~\cite{Babichev:2006cy}, braneworld cosmology~\cite{ramadhan1}, or their gravitational fields and toy models for quantum tunneling in the landscape~\cite{Prasetyo:2015bga}. Surprisingly, the field equations in some of these sub-space of noncanonical defects, known as generalized vortex and monopole, can be reduced into the first-order (BPS)~\cite{Bazeia:2012uc, Casana:2012un}. They are obtained by performing the ordinary Bogomol'nyi trick. 

More surprisingly, these BPS equations can admit exact solutions~\cite{Casana:2014qfa, Casana:2013lna}. This discovery is remarkable because even in the ordinary case it is hard, if not impossible, to obtain such analytic solutions. The BPS monopoles do have exact solutions, the Prasad-Sommerfield solutions~\cite{Prasad:1975kr}, but the attempt to obtain such solutions for BPS vortices has been futile. A closer look will reveal that the appearance of the so-called generalized functions $f(|\phi|)$ and $w(|\phi|)$ ``conspires" to render the scalar and gauge profile functions having analytical solutions.

In this paper, we attempt to present several other exact solutions not covered in~\cite{Casana:2014qfa, Casana:2013lna}. We will largely follow their algorithm. For vortices, we only consider the case of ordinary Mexican-hat potential, while for monopoles we work in the Prasad-Sommerfield limit of the vanishing potential. The paper is organized as follows. In the next section we briefly review the generalized Maxwell-Higgs model and present several exact BPS solutions, both for single and multi vortices. Section III is devoted to the discussion of generalized (non-standard) Yang-Mills-Higgs model and their BPS solutions. Finally, we summarize our results and give some comments in Section IV.

\section{Exact BPS Vortices}

Consider a $(2+1)$-dimensional Lagrangian density studied in~\cite{Bazeia:2012uc}
\begin{equation}
\label{LagVor}
 \mc{L}=-{1\over 4}f^2(|\phi|)F_{\mu\nu}F^{\mu\nu}+w(|\phi|) |D_\mu\phi|^2-V(|\phi|),
\end{equation}
with Minkowski metric $\eta^{\mu\nu}\equiv\text{diag}(+--)$. The BPS condition is achieved when the following self-dual equations are satisfied~\cite{Casana:2014qfa}:
\begin{equation}
\label{eq:selfdualvor}
D_{\pm}\phi=0,\  \ B=\pm\frac{\sqrt{2 V}}{f},
\end{equation}
where $B\equiv\epsilon_{ijk}\partial_{j}A_{k}$ represents the magnetic field.
Employing Nielsen-Olesen ansatz
\begin{equation}
 \phi= v\ g(r) e^{i n\theta},\qquad\qquad \textbf{A}=-{a(r)-n \over e\ r}\hat{\theta},
\end{equation}
the BPS Eqs.~\eqref{eq:selfdualvor} reduce to (after appropriate scaling)
\begin{eqnarray}
\label{BPSvor1}
g'&=&{ag\over r},\\
\label{BPSvor2}
{a'\over r}&=&{g^2-1\over f}.
\end{eqnarray}
The appropriate boundary conditions are
\begin{eqnarray}
\label{bcvor}
g(r=0)&=&0,\ \ \ \ a(r=0)=n,\nonumber\\
g(r\rightarrow\infty)&=&1,\ \ a(r\rightarrow\infty)=0.\nonumber\\
\end{eqnarray}
The function $w$ does not enter into the BPS equations, but it is constrained to satisfy (Eq.(26) in~\cite{Casana:2014qfa})
\begin{equation}
w=-{\left(f\sqrt{2V}\right)'\over 2gg'}.
\end{equation}

The BPS energy can then be written as 
\begin{equation}
E_{BPS}=\mp2\pi v^2\int dr H',
\end{equation}
with 
\begin{equation}
H\equiv af\sqrt{2V}
\end{equation}
an auxiliary function having values 
\begin{eqnarray}
\label{Hvor}
H(0)&\equiv&H_0=\text{finite},\nonumber\\
H(\infty)&\equiv&H_\infty=0,\nonumber\\
\end{eqnarray}
  at the boundaries. The total energy must then be given by 
\begin{equation}
\label{totenvor}
E_{BPS}=2\pi v^2\left|H_0\right|.
\end{equation}
Note that $H_0$ is proportional to winding number $n$. Throughout this paper we limit ourself only to ordinary symmetry-breaking potential, $V(g^2)={1\over 2} \left(1-g^2\right)^2$. 

In~\cite{Casana:2014qfa} several analytical BPS solutions have been presented. Here we try to look for other exact solutions. Our method is the following. First we ``guess" $g$ (or a) that satisfy conditon~\eqref{bcvor}. We substitute this ansatz into Eq.~\eqref{BPSvor1} to obtain $a$ (or g). The remaining equation (Eq.\eqref{BPSvor2}) serves as the condition to determine $f$ and $w$, with which $H$ can be obtained. The acceptable BPS solutions are thus the functions $g(r)$ and $a(r)$ that interpolate between the boundary conditions~\eqref{bcvor} and produce $H(r)$ that satisfies~\eqref{Hvor}. It should be apparent that this method gives the functions $f$ and $w$ in terms of $r$, not explicitly in terms of $|\phi|$. We will show later that some of our exact solutions generalize those found by Casana {\it et al}.

\subsection{$n=1$-BPS Vortices}

Our first guessed function is
\begin{equation}
\label{vorsol1}
g(r)={4\over\pi}\arctan(\tanh(r)).
\end{equation}
This ansatz yields,
\begin{equation}
\label{vorsol1.2}
a(r)={r\ \text{sech(2r)}\over\arctan(\tanh(r))}.
\end{equation}
It is easy to see that both~\eqref{vorsol1}-\eqref{vorsol1.2} satisfy boundary conditions~\eqref{bcvor}. The $f$ function is
\begin{equation}
f(r)={\left[\pi^2r\ \arctan^2(\tanh(r))-16r\ \arctan^4(\tanh(r))\cosh^2(2r)\right]\cosh^2(2r)\over\pi^2\left[r-\arctan{(\tanh(r))}\cosh(2r)+2r\ \arctan(\tanh(r))\sinh(2r)\right]}.
\end{equation}
Here $w(r)$ is not shown since its form is tedious and unilluminating. However we already checked that both functions are positive-definite and regular everywhere, satisfying $f(0)=3/8$, $f(\infty)=0$, and $w(0)=w(\infty)=0$. We also obtain the auxiliary function $H$ as follows
\begin{equation}
H(r)={r^2\ \arctan(\tanh(r))\cosh(2r)\left[\pi^2-16\arctan^2(\tanh(r))\right]\over\pi^2\left[r-\arctan{(\tanh(r))}\cosh(2r)+2r\ \arctan(\tanh(r))\sinh(2r)\right]},
\end{equation} 
with $H_0=3/8$. The energy of this BPS vortex is thus
\begin{equation}
E_{BPS}=0.375E_s,
\end{equation}
where $E_s$ is the total energy of $n=1$ Nielsen-Olesen BPS vortex, $E_s\equiv 2\pi v^2$.\\
\\

We can also try
\begin{equation}
\label{vorsol2}
a(r)=e^{-r^2/2}.
\end{equation}
The Higgs function becomes
\begin{equation}
\label{vorsol2.2}
g(r)=e^{{1\over2}\text{Ei}\left(-{r^2}, 1\right)},
\end{equation}
where $\text{Ei}\left(a, b\right)\equiv\int^{\infty}_{a}{e^{-bu}\over u}du$ is the exponential integral function. They yield
\begin{eqnarray}
f(r)&=&e^{r^2/2}\left(e^{\text{Ei}\left(-r^2/2,\ 1\right)}-1\right),\nonumber\\
w(r)&=&e^{r^2/2}\left(e^{\text{Ei}\left(-r^2/2,\ 1\right)}-1\right)\left[4e^{\text{Ei}\left(-r^2/2,\ 1\right)}+e^{r^2/2}\left(e^{\text{Ei}\left(-r^2/2,\ 1\right)}-1\right)r^2\right],\nonumber\\
H(r)&=&\left(e^{\text{Ei}\left(-r^2/2,\ 1\right)}-1\right)^2.
\end{eqnarray}
From $H_0=1$ we obtain the BPS energy $E_{BPS}=E_s$.\\
\\

Another type of BPS solutions can be obtained by setting
\begin{equation}
\label{vorsol3}
g(r)={e^{{1\over 2\left(r^2+1\right)}}r\over\sqrt{r^2+1}},
\end{equation}
which yields a rather simple gauge function
\begin{equation}
\label{vorsol3.2}
a(r)={1\over\left(r^2+1\right)^2}.
\end{equation}
The corresponding $f$, $w$, and $H$ functions are
\begin{eqnarray}
f(r)&=&1/4\left(r^2+1\right)^2\left[1+r^2-e^{{1\over\left(r^2+1\right)}}r^2\right],\nonumber\\
w(r)&=&1/2\ r^2\left(r^2+1\right)\left[2e^{1\over r^2+1}\left(1+4r^2+6r^4+3r^6\right)-e^{2\over r^2+1}r^2\left(2+3r^2+3r^4\right)-3\left(r^2+1\right)^3\right],\nonumber\\
H(r)&=&{\left[1-\left(e^{1\over r^2+1}-1\right)r^2\right]^2\over 4\left(r^2+1\right)^2}.\nonumber\\
\end{eqnarray}
Here we obtain $H_0=1/4$. The BPS energy is thus
\begin{equation}
E_{BPS}=0.25 E_s.
\end{equation}
This configuration can be generalized by setting
\begin{equation}
\label{vorsol3.1.1}
a(r)={1\over\left(r^2+1\right)^m},
\end{equation}
for arbitrary $m$. This yields a rather complicated Higgs profile
\begin{equation}
\label{vorsol3.1.2}
g(r)=e^{-{\left(1+{1\over r^2}\right)^{m}\ _{2}F_{1}(m, m, 1+m, -1/r^2)\over 2m\left(1+r^2\right)^{m}}},
\end{equation}
where $_{2}F_{1}(q, b, c, z)$ is an ordinary (or Gaussian) hypergeometric function, defined by the following series~\cite{abramowitz}
\begin{equation}
_{2}F_{1}(a,b,c,z)\equiv\sum^{\infty}_{\ell=0}{(a)_{\ell}(b)_{\ell}\over(c)_{\ell}}{z^\ell\over\ell!},
\end{equation}
with $(q)_{\ell}$ the Pochhammer symbol, given by
\begin{equation}
(q)_{\ell}=q(q+1)....(q+\ell-1),
\end{equation}
for $\ell>0$ and $(q)_\ell=1$ for $\ell=0$. One can verify that setting $m=2$ reduces back to BPS configuration~\eqref{vorsol3.1.1}-\eqref{vorsol3.1.2}. The other functions are
\begin{eqnarray}
f(r)&=&{e^{-{\left(1+{1\over r^2}\right)^{m}\ _{2}F_{1}(m, m, 1+m, -1/r^2)\over m\left(1+r^2\right)^{m}}}\left(e^{{\left(1+{1\over r^2}\right)^{m}\ _{2}F_{1}(m, m, 1+m, -1/r^2)\over m\left(1+r^2\right)^{m}}}-1\right)\left(r^2+1\right)^{m+1}\over2m},\nonumber\\
H(r)&=&{\left(1+e^{-{2\left(1+{1\over r^2}\right)^{m}\ _{2}F_{1}(m, m, 1+m, -1/r^2)\over m\left(1+r^2\right)^{m}}}-2e^{-{\left(1+{1\over r^2}\right)^{m}\ _{2}F_{1}(m, m, 1+m, -1/r^2)\over m\left(1+r^2\right)^{m}}}\right)\left(r^2+1\right)\over 2m}.\nonumber\\
\end{eqnarray}
Here we once again avoid showing the explicit form of $w(r)$ due to its over-tediousness. However, we checked that $w(r)$ vanishes for $m=1$. Therefore the allowed values of $m$ for the construction of BPS vortices is $m>1$. The parameter $m$ controls the defect thickness. The greater $m$ the thinner the defect is; {\it i.e.,} the more localized it is. The requirement that $m$ be an integer comes from the fact $H_0=1/2m$ only when $m\in\mathbb{Z}_+$. The BPS energy is thus
\begin{equation}
E_{BPS}=0.5E_s/m.
\end{equation}

\subsection{BPS vortices with $n>1$}

As in~\cite{Casana:2014qfa}, no obvious route in finding exact higher-winding-number BPS vortices seems available. Our attempts suggests that maybe $w(r)$ to be everywhere positive-definite a too stringent constraint to impose. We therefore relax it and impose a weaker condition instead; we allow a small-finite range of $r$ where $w(r)$ is negative but insist that they must satisfy the following conditions:
\begin{equation}
\label{condw1}
\int_0^{\infty} w\ dr>0,
\end{equation}
and
\begin{equation}
\label{condw2}
\int d^2x\ w(|\phi|) |D_i\phi|^2\propto\int_0^{\infty} dr\ r\ w\bigg(g'^2+\left({g\ a\over r}\right)^2\bigg)>0.
\end{equation} 
In this sense the requirement of positive-energy condition still holds.

We start by generalizing solution~\eqref{vorsol2} into
\begin{equation}
a(r)=n\ e^{-r^2/m},
\end{equation}
with $m$ any arbitrary positive real number (not necessarily an integer), $\mathbb{R}_+$. It gives the following Higgs profile function:
\begin{equation}
g(r)=c_1e^{{n\over2}\text{Ei}\left(-{r^2\over m}, 1\right)},
\end{equation}
where $c_1=1$ to satisfy~\eqref{bcvor}. At first this looks like a two-parameter family of BPS vortex solutions, where the vorticity number is controlled by $n$ and the thickness depends on  $m$. However, we will show shortly that for energy consistency $m$ should not be a free parameter but must be related to $n$. Meanwhile, the remaining functions are given by
\begin{eqnarray}
f(r)&=&-{me^{-r^2/m}\left(e^{n\text{Ei}\left({-r^2\over m},1\right)}-1\right)\over 2n},\nonumber\\
H(r)&=&{m\over 2}\left(e^{n\text{Ei}\left({-r^2\over m},1\right)}-1\right)^2.\nonumber\\
\end{eqnarray}
Both are positive-definite monotonically-decreasing functions of $r$ which are rapidly go to zero asymptotically and are regular everywhere. At the core $f(r)$ satisfies $f(0)=n/2m$. On the other hand, $w(r)$ satisfies
\begin{equation}
w(r)={e^{r^2\over m}\left(1-e^{n\text{Ei}\left({-r^2\over m},1\right)}\right)\left[2mne^{n\text{Ei}\left({-r^2\over m},1\right)}+r^2e^{r^2\over m}\left(e^{n\text{Ei}\left({-r^2\over m},1\right)}-1\right)\right]\over n^2}.
\end{equation}
It can be checked that for $n>1$ there exists a finite range of $r$ where $w(r)$ is negative. However, conditions~\eqref{condw1} and \eqref{condw2} still hold. This guarantees that the BPS energy is still bounded from below. Now, since the energy is directly proportional to winding number,
\begin{eqnarray}
E_{BPS}&=&2\pi v^2 \left|H_0\right|\propto n,
\end{eqnarray}
we must set\footnote{In~\cite{Casana:2014qfa} the similar constraint is also discussed, where higher-winding vortex in the context of $\left|\phi\right|^6$-theory is given by $g(r)={r^m\over\left(1+r^n\right)^{{m\over n}}}$ and $a(r)={m\over 1+r^n}$.  In that case $m$ and $n$ must satisfy $n=2m+2$.} 
\begin{equation}
m=2n.
\end{equation}
We then have $H_0=n$. The BPS energy is just equal to the energy of $n$-BPS vortices, 
\begin{equation}
E_{BPS}=2\pi v^2n.
\end{equation}
The BPS solutions are thus
\begin{eqnarray}
g(r)&=&e^{{n\over2}\text{Ei}\left(-{r^2\over 2n}, 1\right)},\nonumber\\
a(r)&=&n\ e^{-r^2/2n},\nonumber\\
f(r)&=&-e^{-r^2/2n}\left(e^{n\text{Ei}\left({-r^2\over 2n},1\right)}-1\right),\nonumber\\
w(r)&=&{e^{r^2\over 2n}\left(1-e^{n\text{Ei}\left({-r^2\over 2n},1\right)}\right)\left[4n^2e^{n\text{Ei}\left({-r^2\over 2n},1\right)}+r^2e^{r^2\over 2n}\left(e^{n\text{Ei}\left({-r^2\over 2n},1\right)}-1\right)\right]\over n^2},\nonumber\\
H(r)&=&n\left(e^{n\text{Ei}\left({-r^2\over 2n},1\right)}-1\right)^2.\nonumber\\
\end{eqnarray}

Another attempt is by generalizing BPS solutions in~\cite{Casana:2014qfa} (Eqs.(38)-(39)):
\begin{eqnarray}
g(r)&=&\left(1-e^{-r^2/n}\right)^{n/2},\nonumber\\
a(r)&=&{r^2\over\left(e^{r^2/n}-1\right)},
\end{eqnarray}
where $n$ labels the vorticity\footnote{As before, we initially set $g(r)=\left(1-e^{-r^2/m}\right)^{n/2}$, where $m$ in general can be arbitrary. This leads to $a(r)={nr^2\over m\left(e^{r^2/m}-1\right)}$. The condition that $E_{BPS}\sim n$ then requires that $m=n$.}. These lead to 
\begin{eqnarray}
f(r)&=&{n\left(e^{r^2/n}-1\right)^2\left[\left(1-e^{-r^2/n}\right)-1\right]\over 2n\left(e^{r^2/n}-1\right)-r^2e^{r^2/n}},\nonumber\\
H(r)&=&{ nr^2\left(e^{r^2/n}-1\right)\left[\left(e^{-r^2/n}-1\right)-1\right]^2\over2n\left(e^{r^2/n}-1\right)-r^2e^{r^2/n}}.
\end{eqnarray} 
Here $H_0=n$. Once again, $w(r)$ is not shown. The BPS energy is simply
\begin{equation}
E_{BPS}=2\pi v^2 n.
\end{equation}

Not every attempt leads to successful result. The conditions~\eqref{condw1} and~\eqref{condw2}, combined with the finiteness of $H_0\neq 0$, severely restrict the availability of our ansatz. For example, we might be tempted to generalize solutions~\eqref{vorsol1}-\eqref{vorsol1.2} by letting
\begin{eqnarray}
g(r)&=&\left({4\over\pi}\right)^n\arctan^n\left(\tanh\left({r\over\sqrt{n}}\right)\right),\nonumber\\
a(r)&=&{\sqrt{n}\ r\ \text{sech}\left({2r\over\sqrt{n}}\right)\over\arctan\left(\tanh\left({r\over\sqrt{n}}\right)\right)}.\nonumber\\
\end{eqnarray} 
This configuration solves the Eqs~\eqref{BPSvor1}-\eqref{BPSvor2} along with conditions~\eqref{bcvor}. As in the previous cases, they satisfy conditions~\eqref{condw1} and \eqref{condw2}. However, careful investigation reveals that they render the energy density, 
\begin{equation}
\varepsilon(r)=\mp{1\over r}{dH\over dr},
\end{equation}
non positive-definite; there exists a finite range of $r$ where $\varepsilon(r)<0$. This negativity of $\varepsilon(r)$ grows with $n$. We decided that this cannot be tolerated since it implies the non-boundedness of energy from below. We therefore discard this solution as non-physical\footnote{Another non-physical solution we found was by generalizing BPS solution (33)-(34) in~\cite{Casana:2014qfa}, 
\begin{eqnarray}
g(r)&=&\tanh^n\left({r\over\sqrt{n}}\right),\nonumber\\
a(r)&=&2\sqrt{n}\ r\ \text{csch}\left({2r\over\sqrt{n}}\right).\nonumber
\end{eqnarray}
This solution, despite satisfying~\eqref{bcvor}, results in the non-positive-definiteness of the energy density. We thus discard it as well.}.

\section{Exact BPS Monopoles}

The analytical solutions for BPS (non-standard) magnetic monopoles was first discussed in~\cite{Casana:2013lna}, whose Lagrangian is given by
\begin{equation}
\label{lagmonop}
\mc{L}=-{1\over 4f\left(\phi^a\phi^a\right)}F_{\mu\nu}^bF^{\mu\mu,b}+{f\left(\phi^a\phi^a\right)\over 2}D_{\mu}\phi^bD^{\mu}\phi^b,
\end{equation}
where $f\left(\phi^a\phi^a\right)$ is a positive-definite ``weight function". Note that we work in the condition $f={1\over w}$. It is easy to show, using the on-shell formalism~\cite{Atmaja:2014fha, Atmaja:2015lia}, that the only possible BPS monopoles in this model can exist should we work in the Prasad-Sommerfield limit ($V\rightarrow 0$)~\cite{Prasad:1975kr}. Assuming the hedgehog ansatz,
\begin{equation}
\phi^a=H(r)\hat{x}^a,\ \ \text{and}\ \ A^a_i=\epsilon_{iak}{\hat{x}^k\over er}\left(W(r)-1\right),
\end{equation}
the Bogomol'nyi equations are ,~\cite{Casana:2012un, Casana:2013lna}
\begin{eqnarray}
\label{BPSmonop}
H'&=&\mp{1-W^2\over er^2f},\\
\label{BPSmonop2}
W'&=&\pm efHW,
\end{eqnarray}
satisfying the following boundary conditions
\begin{eqnarray}
\label{bcmonop}
H(r=0)&=&0,\ \ \ \ W(r=0)=1,\nonumber\\
H(r\rightarrow\infty)&=&\mp1,\ \ W(r\rightarrow\infty)=0.\nonumber\\
\end{eqnarray}
Those equations lead to the following energy density
\begin{equation}
\varepsilon_{BPS}=\mp{1\over er^2}\left[H\left(1-W^2\right)\right]',
\end{equation}
and the corresponding total energy
\begin{equation}
\label{energytotmonop}
E_{BPS}=4\pi\int r^2\varepsilon_{BPS}dr={4\pi\over e}.
\end{equation}

Here we shall present several exact BPS monopole solutions not shown in~\cite{Casana:2013lna}. The ``algorithm" is quite similar to the previous section for BPS vortex: we first deduce $H(r)$ (or $W(r)$) ansatz which is regular everywhere and satisfy boundary conditions~\eqref{bcmonop}, from which we obtain $W(r)$ (or $H(r)$) and $f(r)$.  We have checked that all solutions presented here satisfy $\varepsilon(r)>0$. Due to the regularity and conditions~\eqref{bcmonop} all solutions necessarily have the same total energy~\eqref{energytotmonop}.\\

We start by generalizing solution (19) in~\cite{Casana:2013lna} (for simplicity we set the coupling constant $e=1$):
\begin{equation}
\label{solmonop1}
H(r)={r\over\sqrt{1+r^2}}.
\end{equation}
This results in
\begin{equation}
W(r)={1\over\sqrt{1-r^2e^{2c_1+r^2}}},
\end{equation}
with $c_1$ a constant of integration. To have a real-valued solution, it is clear that we must set $c_1=i\pi/2$; {\it i.e.,}
\begin{equation}
\label{solmonop1.2}
W(r)={1\over\sqrt{1+r^2e^{r^2}}}.
\end{equation}
It then yields
\begin{equation}
f(r)={e^{r^2}\left(1+r^2\right)^{3/2}\over 1+r^2e^{r^2}}.
\end{equation}
It can easily be checked that this function is finite and behaving quadratically around the core, $f(r\rightarrow 0)\approx1+3r^2/2+O(r^4)$, while it goes linearly-divergent for large $r$, $f(r\rightarrow\infty)\approx r$. 
This configuration can be generalized by setting
\begin{equation}
H(r)={r\over\left(1+r^m\right)^{1\over m}}.
\end{equation}
The corresponding profile for $W(r)$ is
\begin{equation}
W(r)={1\over\sqrt{1+r^2e^{2r^m\over m}}}.
\end{equation}
This solution is a valid BPS configuration for all $m$, where $m$ is an arbitrary positive number, $m\in{\mathbb R}_+$. The weight function $f(r)$ is
\begin{equation}
f(r)={\left(1+r^m\right)^{1+{1\over m}}\over r^2}\left({1\over 1+r^2e^{2r^m\over m}}-1\right).
\end{equation}
\\

Another type of solution can be obtained by choosing
\begin{equation}
H(r)={r\over 1+\sqrt{1+r^2}}.
\end{equation} 
We obtain
\begin{equation}
W(r)={2\sqrt{4 r^2 - 
  e^{2 \left(\sqrt{1 + r^2} + c_1\right)} \left(2 + r^2 + 2\sqrt{1 + r^2}\right)}\over\sqrt{
16 r^2 + e^{4 \left(\sqrt{1 + r^2} + c_1\right)} r^2 - 
 8 e^{2 \left(\sqrt{1 + r^2} + c_1\right)} \left(2 + r^2\right)}},
\end{equation}
where once again we must set $c_1=i\pi/2$ to have a real solution satisfying~\eqref{bcmonop}. The corresponding weight function is
\begin{equation}
f(r)={1\over1 - {1\over\sqrt{1 + r^2}}+ e^{-2\sqrt{1 + r^2}}\left(4 + {4\over\sqrt{1 + r^2}}\right)}.
\end{equation}
\\

We can also ``borrow" ansatz from sine-Gordon kink solution for $H(r)$ by setting
\begin{equation}
H(r)={2\over\pi}\arctan(r).
\end{equation}
This rather simple ansatz yields the profile for $W(r)$ as follows
\begin{equation}
W(r)={\left(1+r^2\right)e^{\arctan(r)\over r}\over\sqrt{\left(1+r^2\right)^2e^{2\arctan(r)}+r^2e^{2r\arctan(r)}}}.
\end{equation}
The corresponding weight function is
\begin{equation}
f(r)={\pi\left(1+r^2\right)e^{2r\arctan(r)}\over 2\left(1+r^2\right)^2e^{2\arctan(r)}+2r^2e^{2r\arctan(r)}},
\end{equation}
everywhere regular and positive-definite satisfying $f(0)=\pi/2e^2=0.212584$ and $f(\infty)=1.5708$.\\

The next ansatz we try is as simple as
\begin{equation}
H(r)=1-e^{-r}.
\end{equation}
This $H(r)$-function poduces a-not-so-simple profile for its gauge counterpart
\begin{equation}
W(r)={1\over\sqrt{1+e^{2\left(1-e^{r}+\text{Ei}(r,1)\right)\over r}}}.
\end{equation}
The profile for $f(r)$ is
\begin{equation}
f(r)={e^r\left(1-{1\over 1+e^{2\left(1-e^r+r\text{Ei}(r,1)\right)\over r}}\right)\over r^2}.
\end{equation}
We have checked that $f(r)$ is positive-definite but divergent asymptotically, satisfying $f(0)=e^{2\gamma-2}$, where $\gamma\simeq 0.577216$ is the Euler's constant, and $f(\infty)\rightarrow\infty$. \\

The last variant of solution we consider is
\begin{equation}
H(r)={re^{r}\over\sqrt{1+r^2e^{2r}}},
\end{equation}
which gives
\begin{equation}
W(r)={(1+r)e^{\text{Ei}(2+2r,1)\over e^2}\over\sqrt{r^2e^{e2r}+\left(1+r\right)^2e^{2\text{Ei}(2+2r,1)\over e^2}}}.
\end{equation}
This configuration is possible should we have the following weight function $f(r)$
\begin{equation}
f(r)={e^{e^{2r}-r}\left(1+r^2e^{2r}\right)^{3/2}\over r^2(1+r)e^{e^{2r}}+\left(1+r\right)^3e^{2\text{Ei}(2+2r,1)\over e^2}}.
\end{equation}
This function is finite at the core, $f(0)=e^{1-{2\text{Ei}(2,1)\over e^2}}=0.711083$, and quickly diverges at large $r$.

\section{Discussion and Conclusion}

There are two motivations that guide us in this work. First, we wish to know whether there are other exact BPS solutions for generalized vortices and monopoles not found before. Here, we are able to generalize the results found previously. Sometimes the price to pay is that you might have singular $f(|\phi|(r))$ and $w(|\phi|(r))$ functions at the boundary. However, our profile functions are well-behaved everywhere, interpolating from the core to asymptotic region in a nonsingular way. Moreover, the energy density are regular everywhere. Therefore, we claim that our solutions are indeed legitimate BPS solutions. For the case of BPS monopoles, our solutions all belong to the class of those who do not recover the usual BPS 't Hooft-Polyakov results (according to~\cite{Casana:2013lna}).

Second, we are curious whether the BPS vortices with higher winding number can, in general, be constructed. This result is a little bit more subtle. Finding $n>1$ vortices are by no means an easy task. As stated in~\cite{Casana:2014qfa}, there is no obvious route to follow. They main difficulty in doing so is the condition that the functions $f$ and $w$, as well as the static energy denstiy $\varepsilon$  should be positive-definite. Our BPS monopole results all have positive $f$ but so far suffer from negative values of $w$ in some finite range of $r$. While it is well-known that $w<0$ implies ghost instability we argue, however, that this should not happen since the energy density profile is still positive-definite. This is because, despite our solutions violate the ``strong positive $w$-condition"
\begin{equation}
w(r)>0, \forall r,
\end{equation} 
they still fulfill the ``weaker positive $w$-condition", Eqs.~\eqref{condw1} and~\eqref{condw2}. We therefore claim that they are legitimate BPS solutions. This also brings us to the question of ``can we construct a general formalism to obtain higher-winding BPS vortices once we know the n=1 solutions?". This will be left for our future work.

\acknowledgments
We thank Ardian Atmaja and Dynosio Bazeia for enlightening discussions and suggestions. We also thank the Abdus Salam International Centre for Theoretical Physics (ICTP) in Trieste, Italy, for the hospitality during the completion of this work. This work is supported by the Universitas Indonesia's Research Cluster Grant on ``Non-perturbative phenomena in nuclear astrophysics and cosmology" No.1862/UN.R12/HKP.05.00/ 2015.

\end{document}